\documentclass[final,12pt,nopreprintline]{elsarticle}

\usepackage{graphicx}
\usepackage{float}
\usepackage{amssymb}

\usepackage{lineno}

\usepackage{siunitx}

\usepackage{layouts}

\usepackage{bm}

\journal{Acta Materialia}

\begin{document}

\begin{frontmatter}


\title{Nanoscale lattice strains in self-ion implanted tungsten}



\author{N.W. Phillips\textsuperscript{a*}, H. Yu\textsuperscript{a}, S. Das\textsuperscript{a}, D. Yang\textsuperscript{a}, K. Mizohata\textsuperscript{b},  W. Liu\textsuperscript{c}, R. Xu\textsuperscript{c},  R.J. Harder\textsuperscript{c}, F. Hofmann\textsuperscript{a+}}

\address{\textsuperscript{a} Department of Engineering Science, University of Oxford, Parks Road, Oxford, OX1 3PJ, UK\\
\textsuperscript{b} Accelerator Laboratory, University of Helsinki, P.O. Box 64, 00560 Helsinki, Finland\\ 
\textsuperscript{c} Advanced Photon Source, Argonne National Lab, 9700 S. Cass Avenue, Lemont, IL, USA\\
\textsuperscript{*} Corresponding author E-mail address: nicholas.phillips@eng.ox.ac.uk (N.W. Phillips)\\
\textsuperscript{+} Corresponding author E-mail address: felix.hofmann@eng.ox.ac.uk (F. Hofmann)}

\begin{abstract}
Developing a comprehensive understanding of the modification of material properties by neutron irradiation is important for the design of future fission and fusion power reactors. Self-ion implantation is commonly used to mimic neutron irradiation damage, however an interesting question concerns the effect of ion energy on the resulting damage structures. The reduction in the thickness of the implanted layer as the implantation energy is reduced results in the significant quandary: Does one attempt to match the primary knock-on atom energy produced during neutron irradiation or implant at a much higher energy, such that a thicker damage layer is produced? Here we address this question by measuring the full strain tensor for two ion implantation energies, \SI{2}{\mega\eV} and \SI{20}{\mega\eV} in self-ion implanted tungsten, a critical material for the first wall and divertor of fusion reactors. A comparison of \SI{2}{\mega\eV} and \SI{20}{\mega\eV} implanted samples is shown to result in similar lattice swelling. Multi-reflection Bragg coherent diffractive imaging (MBCDI) shows that implantation induced strain is in fact heterogeneous at the nanoscale, suggesting that there is a non-uniform distribution of defects, an observation that is not fully captured by micro-beam Laue diffraction. At the surface, MBCDI and high-resolution electron back-scattered diffraction (HR-EBSD) strain measurements agree quite well in terms of this clustering/non-uniformity of the strain distribution. However, MBCDI reveals that the heterogeneity at greater depths in the sample is much larger than at the surface. This combination of techniques provides a powerful method for detailed investigation of the microstructural damage caused by ion bombardment, and more generally of strain related phenomena in micro-volumes that are inaccessible via any other technique.  
\end{abstract}

\begin{keyword}
Self-ion implantation \sep neutron irradiation damage \sep Bragg CDI \sep strain tensor \sep defect microscopy 


\end{keyword}

\end{frontmatter}


\section{Introduction}
\label{S:Intorduction}
Tungsten (W) is an important candidate material for future fusion power, where it is the material of choice for the plasma-facing armour \cite{Rieth2013,Knaster2016}. These components will be subjected to extreme environments in-service. Typical operating temperatures are expected to be in the region of \SIrange[range-phrase = --, range-units = single]{570}{1270}{\kelvin} \cite{Knaster2016}, with heat loads of \SIrange[range-phrase = --, range-units = single]{2}{20}{\mega\watt/\m^2} and a neutron flux expected to result in the accumulation of several displacements per atom (dpa)/year \cite{Gilbert2012}. Tungsten's high melting point (\SI{3422}{\celsius}), good thermal conductivity, low sputtering rate and low tritium retention make it a suitable candidate material for the first wall amour and divertor \cite{Knaster2016}. However, a significant amount of damage will accumulate in these environments. Beside neutron-induced cascade damage, transmutation will introduce foreign elements such as rhenium, osmium, tantalum, hydrogen and helium. Further hydrogen and helium will diffuse from the plasma itself \cite{Gilbert2012}. These processes have been shown to alter the material properties, resulting in hardening, embrittlement, creep and lattice swelling \cite{Hardie2013,Rieth2013,Knaster2016,Yi2016,Das2018b,Das2019}. Resultant residual stresses can act as offsets to other fatigue loading, potentially reducing fatigue life, which is likely to be the limiting factor for plant service life. The combined effects must be better understood in order to develop ways to inhibit these changes in the design of power generating plants and to estimate the lifetime of functional plasma facing components. \cite{Knaster2016,Gilbert2012,Rieth2013,Das2019}. 

To better understand neutron irradiation damage processes, self-ion implantation can be used to mimic fusion neutron-induced irradiation damage. Self-ion analogues are often selected in favour of neutron irradiation as ion fluxes are much higher, enabling the desired dpa level to be obtained in minutes or hours rather than months or years. Furthermore, the neutron spectrum of typical fission sources fails to replicate the high energy neutrons (\SI{14.1}{\mega\eV}) that will be produced by the DEMO fusion reactor \cite{Gilbert2015,Knaster2016}. Neutron irradiation also results in some sample radioactivity, often necessitating storage while the material cools before analysis can continue. 

The impact of energetic ions and the subsequent cascades within tungsten have been investigated both computationally \cite{Sand2014, Sand2017} and via laboratory-based experiments \cite{Yi2016,Mason2018,Yu2018}. However, information better representative of the irradiation that will occur in a fusion reactor is still required. This includes, but is not limited to: primary knock-on atom (PKA) energy and rate dependence, cascade interactions at the mesoscale and the heterogeneity of damage microstructure. Sand \textit{et al.} \cite{Sand2013, Sand2014, Sand2017}, provide a foundation for such studies via molecular dynamics simulations, showing the scale and distribution of defect structures within a cascade including the formation of $1/2\langle111\rangle$ and $\langle100\rangle$ interstitial and vacancy type dislocation loops for \SIrange[range-phrase = --, range-units = single]{100}{400}{\kilo\eV} self-ion implanted tungsten at \SIrange[range-phrase = --, range-units = single]{0}{4}{\kelvin}. This provides a fairly complete understanding of individual impacts of energetic particles. By beginning molecular dynamics simulations with pre-damaged iron and tungsten systems, Sand \textit{et al.} have shown that cascade overlap, as well as cascade splitting, are significant factors resulting in heterogeneity of the damage structure \cite{Sand2018}.\textit{In situ} observation of self-ion implantation in tungsten transmission electron microscopy (TEM) foils by Yi \textit{et al.} \cite{Yi2016}, provides further confirmation of the types and distributions of defects produced as the microstructure evolves \cite{Zhou2007}.\footnote{For loops visible at a specified diffraction condition. Loop sizes below \SI{1.5}{\nano\metre} remain invisible.} More recently, developments in stereo-imaging TEM by Yu \textit{et al.} \cite{Yu2018} have made it possible to extract the three-dimensional position of irradiation induced dislocation loops in a 40 nm thick foil for low damage levels (0.01 dpa). In addition to being restricted to thin foils, three-dimensional imaging becomes problematic for higher damage levels where defects overlap, and thus far no TEM tomography data for this regime has appeared in the literature.   

The aforementioned works use a low ion energy (around \SI{150}{\kilo\eV}) as it is most representative of the PKA energy generated by a fusion neutron collision \cite{Sand2014}. This eases the computational requirement for simulation as the required volume remains relatively small. However, performing laboratory-based measurements on low energy implanted samples presents a significant challenge. This is primarily due to two effects. Firstly, the implanted layer from \SI{150}{\kilo\eV} W\textsuperscript{+} in tungsten is only tens of nanometres thick, making the detection of signal from such a small volume difficult to separate from that of the unimplanted bulk. Furthermore, commonly used sample preparation methods, such as Focused Ion Beam (FIB) liftout procedures for producing transmission electron microscopy samples, risk modifying the surface layer \cite{Hofmann2017, Hofmann2018}. Secondly, free surface effects begin to play a significant role in such a thin damage layer, resulting in a deviation from the behaviour expected in a bulk sample for a similar amount of damage \cite{Yu2018,Mason2014,Das2019}. This has led to the general approach of using significantly higher ion energies for measurements in order to circumvent the complications associated with low ion energy implantation (e.g. for electron microscopy \cite{Ferroni2015,Ciupinski2013}, nano-indentation \cite{Beck2017,Das2018b}, and thermal transport measurements \cite{Hofmann2015a}). 

Das \textit{et al.} \cite{Das2018b,Das2018} have previously used 0.05 to 1.8 \SI{}{\mega\eV} He\textsuperscript{+} implantation in tungsten to generate a sufficiently thick (\SIrange[range-phrase = --, range-units = single]{2}{3}{\micro\metre}) implantation layer for investigation by differential aperture X-ray microscopy (DAXM), which allows depth-resolved measurements of lattice strain with sub-micron spatial resolution \cite{Larson2002,Liu2004}. In the case of heaver ions such as tungsten, the limited depth resolution (\SIrange[range-phrase = --, range-units = single]{0.5}{1}{\micro\metre}) of DAXM necessitates the concession of using implantation at energies of around \SI{20}{\mega\eV} to ensure a clearly distinguishable implantation layer is produced to a depth of \SIrange[range-phrase = --, range-units = single]{2}{3}{\micro\metre}. In tungsten, above ion energies of a few hundred \SI{}{\kilo\electronvolt}, it is known that rather than producing a single cascade, splitting into a  number of sub-cascades with a PKA energy in the range of \SIrange[range-phrase = --, range-units = single]{150}{300}{\kilo\electronvolt} occurs \cite{Sand2017,Sand2018}. The potential for cascade overlap and higher local energy dissipation may lead to a defect evolution distinctly different to that anticipated from \SI{14.1}{\mega\eV} neutrons. As such it is vital to assess whether ion-implantation damage obtained using different implantation energies is comparable simply on the basis of dpa.  

Bragg coherent diffractive imaging (BCDI) is uniquely suited to the investigation of moderate energy ion implantation, as it has the combination of nanometre spatial resolution in three dimensions, high strain sensitivity and does not requiring thinning of the sample to tens of nanometres thickness. Data collection is performed by illuminating the sample with a coherent, monochromatic X-ray beam whilst satisfying the Bragg condition for a given reflection (\textit{hkl}). The sample is then rotated through the Bragg condition, and an over-sampled diffraction pattern is collected in the far field. Once a three-dimensional reciprocal space map of the intensity surrounding a Bragg peak has been collected, phase retrieval algorithms need to be used to recover the complex-valued, real space sample function, where the amplitude is proportional to the electron density and the phase is proportional to the lattice displacement along the direction of the scattering vector \cite{Robinson2001,Williams2003}. The sample size is  restricted to sub-micron crystals owing to the sub-micron coherence lengths in the hard X-ray regime at current synchrotron sources. Until recently, BCDI has been limited to materials that form isolated, micro-crystals \cite{Pfeifer2006,Newton2010,Clark2015,Ulvestad2017c} and a few samples which can be annealed and retain sub-micron grain size \cite{Vaxelaire2014,Yau2017}. By measuring a multi-reflection BCDI (MBCDI) dataset consisting of three or more non-parallel reflections, the full lattice strain tensor can be determined \cite{Newton2010,Hofmann2017}. Applying our recently demonstrated protocol for top-down fabrication of BCDI strain microscopy samples \cite{Hofmann2020}, we demonstrate that using MBCDI, the strain associated with ion implantation can be investigated over previously inaccessible length scales in the range of tens of nanometres up to a micrometre. 

Here we compare strain in \SI{20}{\mega\eV} self-ion implanted tungsten, measured using DAXM, to the strain distribution in \SI{2}{\mega\eV} self-ion implanted tungsten measured using MBCDI and high-resolution electron back-scatter diffraction (HR-EBSD). This seeks to address the central question: Whether lattice strain is simply a function of dpa, or whether the ion-energy also plays an important role in defect structure formation and how this structure varies throughout the implanted layer. BCDI, provides more than an order of magnitude improvement in volumetric spatial resolution compared to state-of-the-art DAXM. This enables us to directly probe the full strain field associated with the sub-micron implantation layer and makes it possible to resolve the spatial heterogeneity of the implantation induced strain and damage microstructure in three-dimensions.

\section{Experimental}
\label{S:experimental}
\subsection{Sample preparation}
\label{SS:sam_prep}
As rolled tungsten (99.99~\% purity) was mechanically ground and then polished using a gradient of diamond paste. \SI{0.1}{\micro\m} colloidal silica was used to obtain a high quality surface. The material was annealed and fully recrystallised in a vacuum furnace at \SI{1500}{\celsius} for 20 hours (\SI{2}{\mega\eV}) and 10 hours (\SI{20}{\mega\eV}) using a heating and cooling rate of \SI{4}{\celsius/\min}. Parallel studies showed an annealing time of 10 hours to be sufficient. Grains with the (010) lattice plane aligned parallel to the sample surface, presented a flat surface after annealing (see Figure~\ref{F:F1_method}(c)). Some facetting of the surface within differently orientated grains was seen \cite{Yu2019}. This did not impact the quality of the preparation as only the (010) surface normal orientation was investigated.

\subsection{Ion implantation}
\label{SS:ion_imp}
Implantation of samples was performed at the University of Helsinki using a raster scanned beam at room temperature. One sample was implanted with \SI{2}{\mega\eV} W\textsuperscript{+} ions and the second sample with \SI{20}{\mega\eV} W\textsuperscript{+} ions. The ion fluence was \SI{1.02e13}{ions/\cm^2} for the \SI{2}{\mega\eV} sample and \SI{2.53e13}{ions/\cm^2} for the \SI{20}{\mega\eV} sample. Owing to the reduced scattering cross-section at higher energies, an increased ion fluence is required for the \SI{20}{\mega\eV} implantation in order to achieve a comparable dpa. This gave a peak dpa of approximately \SI{0.07}{dpa} and \SI{0.1}{dpa} for the \SI{2}{\mega\eV} and  \SI{20}{\mega\eV} samples respectively. The dpa and implanted self-ion concentration as a function of depth was estimated using ten thousand ion trajectories simulated in SRIM \cite{Ziegler2010} via the quick K-P method, using a displacement threshold of \SI{68}{\electronvolt} \cite{ASTM2009}. The anticipated depth profiles are shown in Figure~\ref{F:F1_method}(a-b). The slightly lower damage dose in the \SI{2}{\mega\eV} sample may result in a less heterogeneous nano-scale damage microstructure. Unfortunately, the direct comparison with the \SI{20}{\mega\eV} implanted sample cannot be easily made, since the micro-beam Laue diffraction technique used to study the \SI{20}{\mega\eV} sample lacks the spatial resolution required to resolve nano-scale strain heterogeneity. Previous mechanical property \cite{Das2019a} and thermal transport studies \cite{Reza2020}, performed on \SI{20}{\mega\eV} self-ion implanted tungsten as a function of damage dose, suggest that there is little difference in the damage microstructure of \SI{0.07}{dpa} and \SI{0.1}{dpa} damaged tungsten.   

\subsection{Liftout procedure for \SI{2}{\mega\eV} implanted MBCDI sample}
\label{SS:BCDI_liftout}
\begin{figure}[H]
\centering\includegraphics[width=1\textwidth]{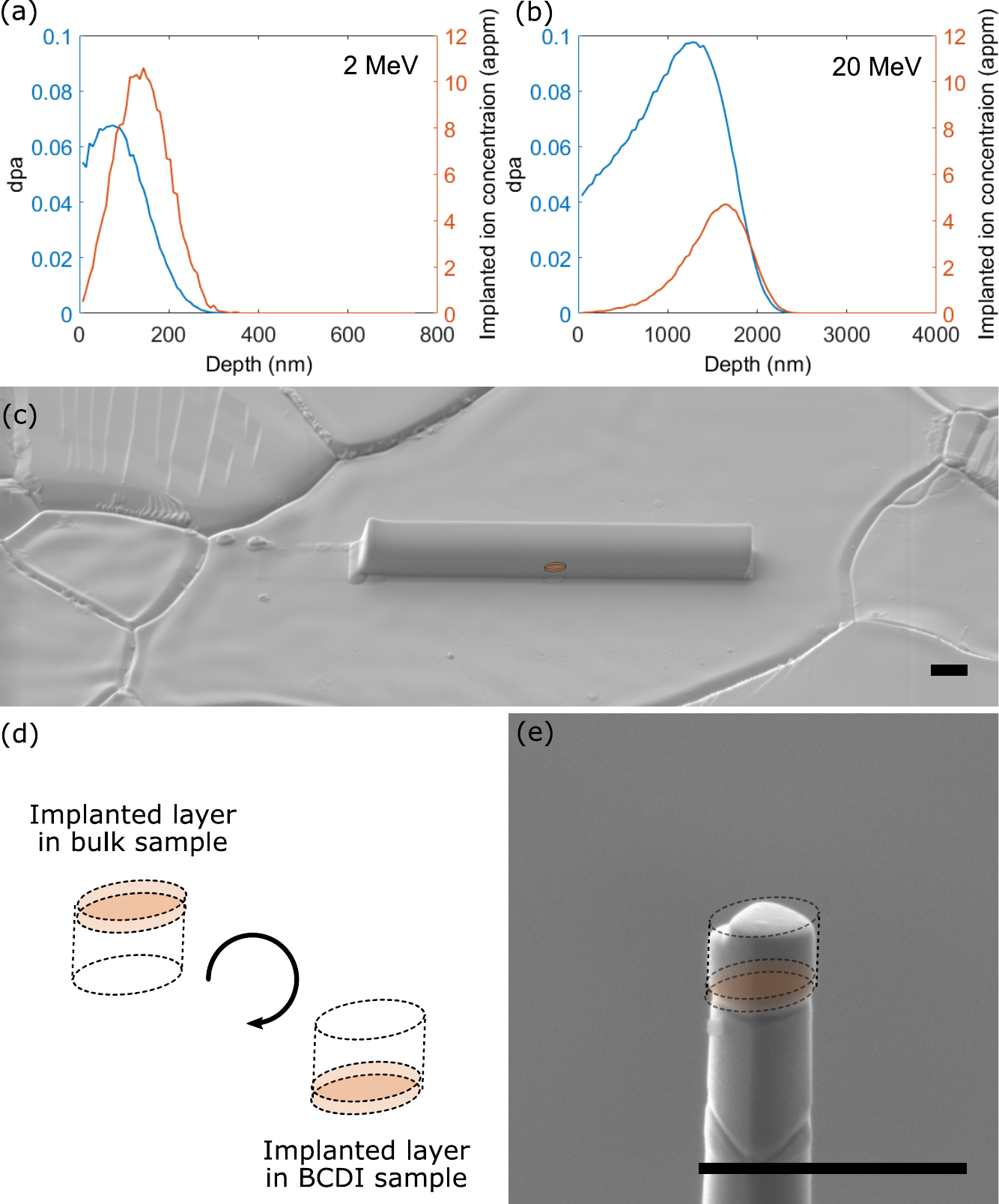}
\caption{Plot of the expected dpa and injected ion concentration as a function of depth after \SI{2}{\mega\eV} (a) and \SI{20}{\mega\eV} (b) self-ion implantation. (c) SEM micrograph of the sample surface for the (010) orientated grain that was used to fabricate the MBCDI strain microscopy sample from the \SI{2}{\mega\eV} ion-damaged specimen. Here the initial rectangular shaped platinum cap marks the lift-out region and the approximate position of the BCDI sample is shown. (d) Schematic of the MBCDI strain microscopy liftout, highlighting the initial upwards facing implanted surface layer and its final position at the bottom of the strain microscopy volume. (e) SEM image of the finished sample with the location of the implanted layer superimposed in orange. The scale bar for the micrographs (c) and (e) is \SI{2}{\micro\metre}.}
\label{F:F1_method}
\end{figure}

To produce a sufficiently small strain microscopy sample for BCDI, a sub-micron volume was extracted from the near-surface material of the \SI{2}{\mega\eV} implanted sample using FIB milling (Figure~\ref{F:F1_method}). This followed our previously developed protocol \cite{Hofmann2020}: Using EBSD, a (010) orientated grain was selected and a \SI{300}{\nano\metre} thick protective cap deposited over the implanted surface by electron beam assisted deposition of platinum. The thickness of the cap was increased to \SI{4}{\micro\metre} using gallium-FIB assisted deposition of platinum (shown in Figure~\ref{F:F1_method}(c)). This ensures that the self-ion implanted surface is never exposed to energetic gallium ions which have been shown to result in strains extending over hundreds of nanometers into the material \cite{Hofmann2017,Hofmann2018}. A liftout, similar to that used when preparing a TEM or atom probe tomography sample \cite{Prosa2009}, was prepared on a Zeiss Auriga dual beam FIB/SEM. This is used to extract a $25\times{}2\times{}6~$\SI{}{\micro\metre}\textsuperscript{3} $(L\times{}W\times{}H)$ lamella, which is then turned upside down (Figure~\ref{F:F1_method}(d)) and attached to a \SI{2}{\micro\metre} diameter silicon pillar using platinum deposition (Figure~\ref{F:F1_method}(e)). This results in the implanted layer being at the bottom of the BCDI sample volume. The sample is then milled to a size of approximately \SI{1}{\micro\metre} in each dimension. Finally, low energy, \SI{2}{\kilo\eV}, Ga\textsuperscript{+} polishing was used to clean off the damage from previous FIB milling, removing ~\SI{\sim{100}}{\nano\metre} of material from each side of the sample. This eliminates most of the damage from previous FIB milling steps \cite{McCaffrey2001,Thompson2007,Hofmann2017,Hofmann2018}.  

\subsection{MBCDI measurement and analysis}
\label{SS:BCDI_analysis}
White-beam micro-Laue diffraction was used to determine the crystallographic orientation of the prepared BCDI sample prior to collection of MBCDI data. This was performed following our previously developed protocol \cite{Hofmann2017a}, at 34-ID-E of the Advanced Photon Source (APS) using LaueGo \cite{Tischler2010} for the analysis of the Laue diffraction data. The \textbf{UB} matrix \cite{Busing1967} is used to provide an accurate description of the sample orientation on a kinematic mount, which is in-turn used to mount and pre-align the sample for MBCDI measurements.   

MBCDI data were collected at beamline 34-ID-C of the APS using a monochromatic X-ray beam of \SI{10}{\kilo\eV}, focused by Kirkpatrick-Baez mirrors to a FWHM of \SI{1.1 x 1}{\micro\metre} (horizontal~$\times{}$~vertical). An MBCDI dataset consisting of five reflections, ($\bar{1}$01), (011), (110), (1$\bar{1}$0) and (0$\bar{1}$1), was measured as the sample was rocked through the Bragg condition forming a three-dimensional, over-sampled, reciprocal space map for each reflection. An angular range of \SI{0.6}{\degree} using \SI{0.0025}{\degree} increments was used. An ASI Timepix detector with 256~$\times{}$~256, \SI{55}{\micro\metre} square pixels and a GaAs sensor was used to record the diffraction patterns at a detector distance of \SI{1}{\metre}. $30\times{}$\SI{0.05}{\s} exposures were taken at each point, with the measurement for each reflection repeated 10 times, maximising the intensity between each scan by centring the sample in the horizontal and vertical directions with respect to the beam and performing the subsequent scan centred about the rocking angle corresponding to the peak intensity. This minimises the effect of long timescale drift on the data quality. For a comprehensive description of the experimental geometry see \cite{Yang2019}. Diffraction patterns were flat-field and dead-time corrected, and then aligned to minimise their Pearson correlation coefficients using a 3D version of the algorithm described by Guizar-Sicairos \textit{et al.} \cite{Guizar-Sicairos2008a}. Those satisfying the threshold criterion of a correlation coefficient exceeding 0.975 were summed for phase retrieval analysis. Ten scans were summed for the ($\bar{1}$01), (011), (110) and (1$\bar{1}$0) reflections, whilst nine scans were summed for the (0$\bar{1}$1) reflection.

Phasing of BCDI data employed previously established approaches for phase retrieval \cite{Clark2012,Clark2015c} and was performed over 4 successive cycles, where the previous result was used to seed the subsequent cycle (with the exception of the first cycle, which was initialised using a random phase guess). Details of the phasing procedure are given in Supplementary Information Section 1.

After applying the appropriate coordinate transform to the retrieved data \cite{Yang2019}, the projection of the crystal electron density is given by the absolute value of the recovered data whilst the phase is proportional to the lattice displacement \textbf{u}(\textbf{r}), along the direction of the Bragg vector \textbf{q}$_\textit{hkl}$, \textit{i.e.} $\phi{}(\textbf{r})=\textbf{q}_\textit{hkl}\cdot{}\textbf{u}(\textbf{r})$. Employing our recently developed method of accounting for phase wraps and phase discontinuity due to crystal defects  \cite{Hofmann2020}, the strain is computed as follows:

Firstly, phase offsets of $\pm{\pi/2}$ (two are typically sufficient) are applied to the recovered phase for each reflection in the detector conjugated coordinate frame and the values constrained between \num{0} and \num{2\pi{}}. Then the recovered phase and the two copies are transformed into an orthogonal sample space \cite{Yang2019} common to all reflections in the MBCDI dataset. The spatial derivatives of $\phi{}(\textbf{r})$ are then computed and the pixel-by-pixel gradient with the smallest magnitude used for computing the lattice strain tensor through the minimisation of the squared error of the displacement gradient and the gradient of the measured phase for each reflection, 
\begin{equation}\label{eq:1}
E(\textbf{r})_i=\sum_{hkl,j}\left(\textbf{q}_{hkl}\cdot{}\frac{\partial \textbf{u}(\textbf{r})}{\partial j} - \frac{\partial \phi(\textbf{r})}{\partial j} _{hkl}\right)^{2},
\end{equation}
where \textit{j} refers to the spatial x, y or z coordinate. 

For a complete discussion of this approach the reader should refer to \cite{Hofmann2020}. The lattice strain tensor components and lattice rotations are then given by 
\begin{equation}\label{eq:2}
\bm{\varepsilon}(\textbf{r}) = \frac{1}{2}\left(\text{grad}\: \textbf{u}(\textbf{r}) + \left[\text{grad}\: \textbf{u}(\textbf{r})\right]^\mathrm{T}\right),
\end{equation}
\begin{equation}\label{eq:3}
\bm{\omega}(\textbf{r}) = \frac{1}{2}\left(\text{grad}\: \textbf{u}(\textbf{r}) - \left[\text{grad}\: \textbf{u}(\textbf{r})\right]^\mathrm{T}\right).    
\end{equation}

For convenience of displaying the results, the $\bm{\varepsilon}(\textbf{r})$ and $\bm{\omega}(\textbf{r})$ tensors calculated by Equations \ref{eq:2} and \ref{eq:3} are displayed in a $3 \times 3$ map format as shown in Table \ref{tab:strain_rot_tab}

\begin{table}[h]
    \centering
    $\left[
    \begin{tabular}{c c c}
    $\varepsilon_{xx}=\frac{\partial \mathrm{u}_x}{\partial x}$ & $\varepsilon_{xy}=\frac{1}{2}\left(\frac{\partial \mathrm{u}_x}{\partial y} + \frac{\partial \mathrm{u}_y}{\partial x}\right)$ & $\varepsilon_{xz}=\frac{1}{2}\left(\frac{\partial \mathrm{u}_x}{\partial z} + \frac{\partial\mathrm{u}_z}{\partial x}\right)$ \\
    \cline{1-1}\multicolumn{1}{c!{\vrule width 1pt}}{$\omega_{z}=\frac{1}{2}\left(\frac{\partial \mathrm{u}_y}{\partial x} - \frac{\partial \mathrm{u}_x}{\partial y}\right)$} & $\varepsilon_{yy}=\frac{\partial \mathrm{u}_y}{\partial y}$ & $\varepsilon_{yz}=\frac{1}{2}\left(\frac{\partial \mathrm{u}_y}{\partial z} + \frac{\partial \mathrm{u}_z}{\partial y}\right)$ \\
    \cline{2-2}{$\omega_{y}=\frac{1}{2}\left(\frac{\partial \mathrm{u}_x}{\partial z} - \frac{\partial \mathrm{u}_z}{\partial x}\right)$} & \multicolumn{1}{c!{\vrule width 1pt}}{$\omega_{x}=\frac{1}{2}\left(\frac{\partial \mathrm{u}_z}{\partial y} - \frac{\partial \mathrm{u}_y}{\partial z}\right)$} & $\varepsilon_{zz}=\frac{\partial \mathrm{u}_z}{\partial z}$ \\
    \end{tabular}
    \right]$
    \caption{Lattice strain tensor components and lattice rotations. The representation of lattice strain and rotation in Figures \ref{F:F4_StrainRotYZ}, \ref{F:F5_StrainRotZX} and \ref{F:F7_HREBSDvsBCDIab} follow this layout.}
    \label{tab:strain_rot_tab}
\end{table}

Previous studies used a single voxel reference point within the sample from which an offset for either the phase or its gradient is computed and applied to the entire sample volume \cite{Hofmann2017,Hofmann2017a,Hofmann2020}. In this work we improve the robustness of the normalisation by normalising the average gradient within a 40 pixel cube of data centred in the unimplanted region of the sample where we expect the average strain to be small. With precise knowledge of the diffractometer angles it would be possible to perform a final correction in the form of an offset of the absolute strain in the sample, removing the need for there to be a strain free reference point within the sample volume. In practice, achieving the required level of calibration would be very challenging as the absolute strain resolution of BCDI is approximately $2\times 10^{-4}$ \cite{Hofmann2020}. This is especially the case for MBCDI, where both the detector and sample are moved significantly over the course of the measurement and misalignment between rotation centres is often present to some degree. Fortunately, in the present sample it is not necessary to have an absolute measurement as the unimplanted region may be used as an in-built reference.

\subsection{X-ray micro-beam Laue diffraction measurements}
\label{SS:BCDImeasurement}
X-ray micro-beam Laue diffraction was used to measure the lattice parameter within the implanted layer of the sample exposed to \SI{20}{\mega\eV} self-ions and the unimplanted bulk material beneath. This is possible through DAXM \cite{Larson2002} and was performed at beamline 34-ID-E of the APS. The differential aperture, a platinum knife-edge, is scanned across the surface of the sample in sub-micrometre steps, enabling the depth from which the signal originates to be determined via a ray tracing approach \cite{Larson2002,Liu2004}. DAXM was performed for a range of monochromatic energies, which provides the lattice spacing as a function of depth within the material. An energy-wire scan (EW-DAXM) was performed, collecting the scattered intensity for a single (060) out-of-plane reflection. The spatial resolution of DAXM is closely linked to the size of the probe used to illuminate the sample. Here the probe has a full-width-at-half-maximum (FWHM) of \SI{300}{\nano\metre} in both the horizontal and vertical directions with depth resolution of approximately \SI{500}{\nano\metre}. Strain sensitivity of \num{1e-4} is routinely achieved using this technique. Further details are provided elsewhere \cite{Liu2004,Hofmann2013}. In total, three sample positions were analysed on the umimplanted reference sample and four positions on the \SI{20}{\mega\eV} self-ion implanted \SI{0.1}{dpa} sample. Data analysis was performed using LaueGo \cite{Tischler2010}. 

\subsection{HR-EBSD measurements}
\label{SS:EBSD}
HR-EBSD was used to obtain high resolution maps of the deviatoric strain tensor and lattice rotation at the surface of the \SI{2}{\mega\eV} self-ion implanted tungsten sample. The data were collected on a Zeiss Merlin scanning electron microscope (SEM) equipped with a Bruker eFlash detector, using an accelerating voltage of \SI{20}{\kilo\eV}, a current of \SI{15}{\nano\A}, with an EBSD pattern size of 800 $\times{}$ 600 pixels and a step size of \SI{16.7}{\nano\meter}. Employing a cross-correlation based analysis approach for EBSD provides strain sensitivity as high as $2 \times 10^{-4}$ \cite{Wilkinson1996}. The analysis was performed using the MATLAB-based HR-EBSD software developed by Britton and Wilkinson \cite{Britton2012}.

In contrast to the aforementioned X-ray diffraction methods, it is worth noting that HR-EBSD is insensitive to volumetric strains and that the signal originates from a sample volume within the first few nanometers of the surface \cite{Zaefferer2007}.

The raw data and analysis code may be made available upon request. MBCDI analysis code is available at \cite{felix2019a}.

\section{Results}
\label{S:Results}

\subsection{X-ray micro-beam Laue diffraction results}
\label{SS:Laue_results}
\begin{figure}[H]
\centering\includegraphics[width=.5\textwidth]{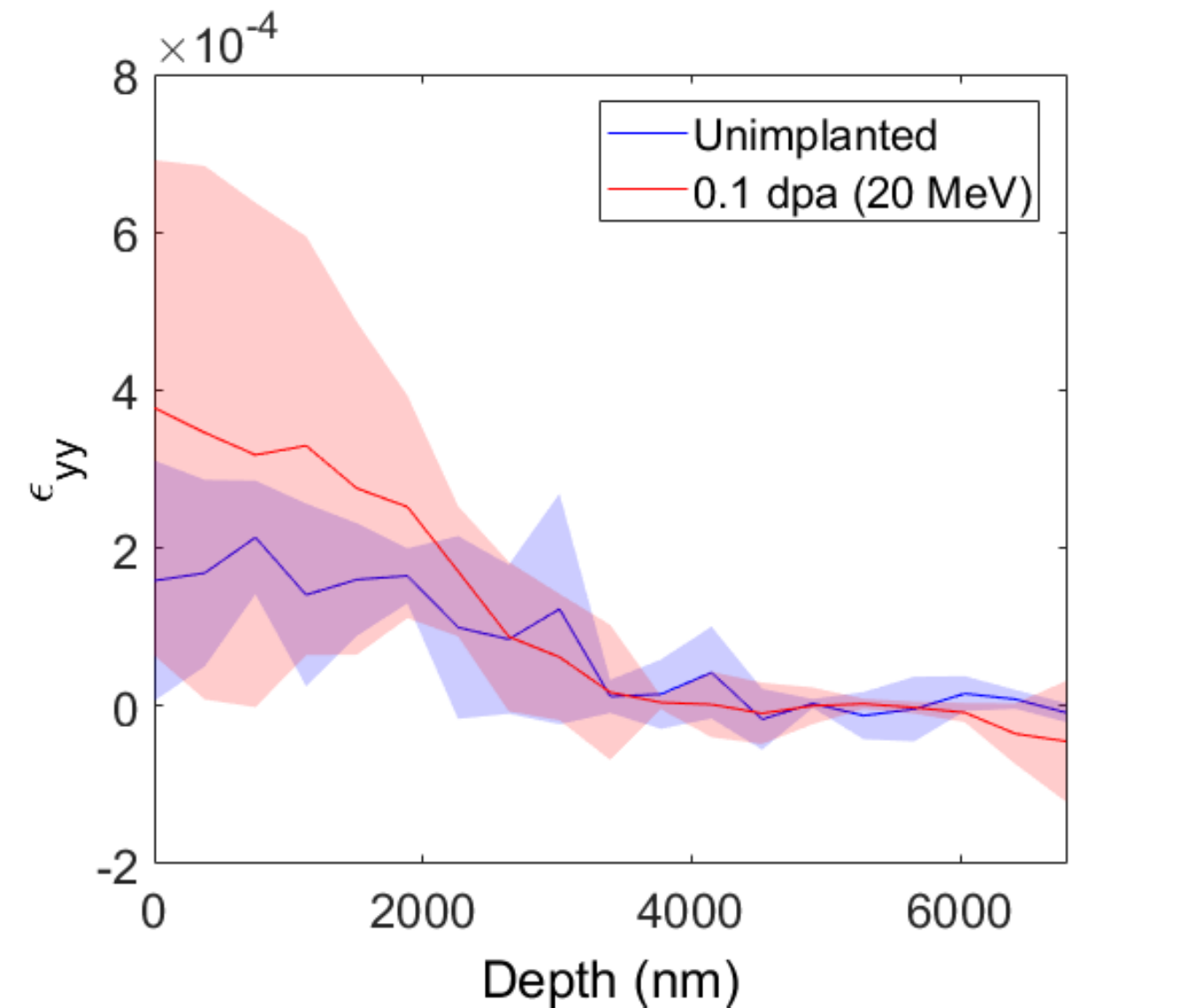}
\caption{DAXM strain profile for \SI{20}{\mega\eV} implantation as a function of depth. Shown is the averaged $\bm{\varepsilon}{_{yy}}$ (out-of-plane) strain component from a series of sample positions, unimplanted (blue) and 0.1 dpa (red). The shaded bounds indicate one standard deviation for the measured data. The implanted layer extends approximately \SI{2000}{\nano\metre} from the surface and has an average strain of \num{3.5e-4} within the first \SI{1000}{\nano\metre} of the surface. As expected, the strain is close to zero for depths below the implantation layer and in the unimplanted reference sample.} 
\label{F:F2_LaueStrain}
\end{figure}

Previous studies employing EW-DAXM to investigate He\textsuperscript{+} implantation in tungsten reported that the in-plane strain components are close to zero \cite{Hofmann2015b,Das2018,Das2018b}. Hence, we only measure the depth variation of the out-of-plane strain component. To exclude any systematic dependence on grain orientation, we only consider grains with (010) surface normal. EW-DAXM depth-strain profiles from \SI{20}{\mega\eV} W\textsuperscript{+} self-ion implantation are displayed in Figure \ref{F:F2_LaueStrain}. These show a positive strain, \textit{i.e.} an expansion of the lattice, to a depth of \SI{2}{\micro\metre}, which is in agreement with the expected thickness for the implanted layer of approximately \SI{2}{\micro\metre}, as calculated by SRIM \cite{Ziegler2010}, shown in (Figure~\ref{F:F1_method}(a-b)). In tungsten the relaxation volume associated with mono-vacancies is small and negative (V\textsubscript{1} = -0.37 $\Omega$, see Hofmann \textit{et al.} \cite{Hofmann2015b} for further vacancy configurations), whilst that of a self-interstitial atom (SIA) is larger and positive (SIA = 1.68 $\Omega$) \cite{Hofmann2015b}. As vacancies in tungsten are immobile at room temperature, whilst SIAs are highly mobile \cite{Hofmann2015b}, there are at most, similar numbers of vacancies and SIAs within the material. Overall this results in the observation of lattice swelling as the relaxation volume of a Frenkel pair, \textit{i.e.} a vacancy and SIA pair, is positive. Both the nominally unimplanted bulk and the unimplanted reference sample show little strain, as is expected for tungsten after annealing at 1500$^\circ$C. The small amount of strain observed near the surface of the unimplanted sample is attributed to damage introduced during sample polishing, an observation supported by the presence of small visible scratches on the sample surface. DAXM measurements on purposely introduced scratches in nickel showed that the strain fields associated with scratches can extend to depths tens of times larger than the scratch depth \cite{SuominenFuller2007}. It should be noted that even for specialised micro-beam Laue instruments, these measurements are approaching the limit of both depth and strain sensitivity, hence the large standard deviation for this data is not unexpected \cite{Hofmann2011,Hofmann2013}. The increase in standard deviation within the implanted layer perhaps hints at spatial heterogeneity caused by the clustering of defects within the implanted layer. This demonstrates the need for the development of methods, such as BCDI, that address these resolution limits.  

\subsection{Multi-Reflection Bragg Coherent Diffractive Imaging}
\label{SS:BCDI_results}
\begin{figure}[H]
\centering\includegraphics[width=.5\textwidth]{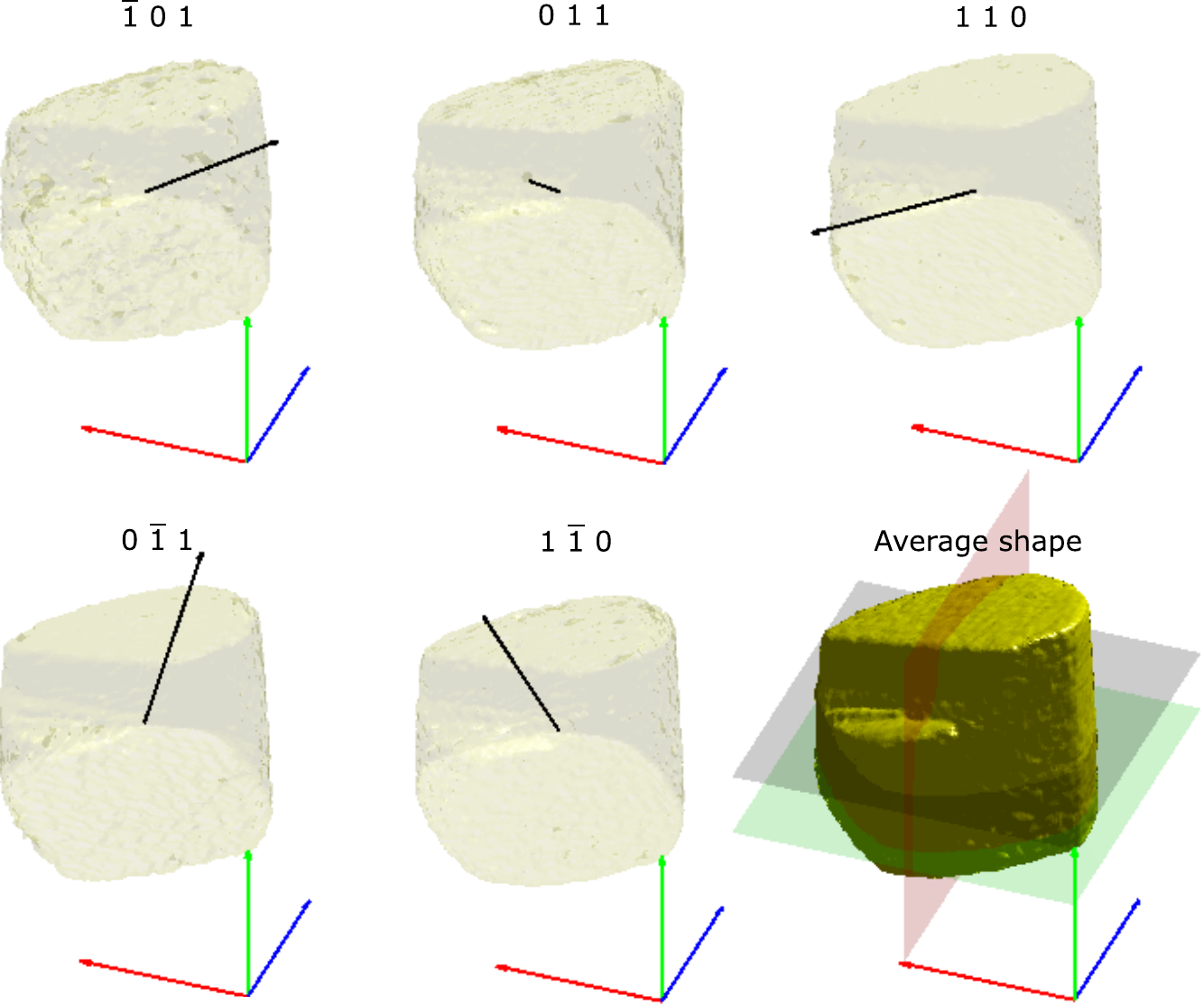}
\caption{Rendering of the sample morphology for each measured reflection labelled with the scattering vector \textit{hkl}. The Bragg vector direction for each reflection is indicated by the black arrows. The averaged strain microscopy sample morphology from all five measured reflections is shown in the lower right. The \SI{500}{\nano\meter} long arrows, red, green and blue comprise a right-handed coordinate system designating the x, y and z axis respectively, which later correspond to the strain tensor components. The red y-z (vertical) and green x-y (horizontal implanted), black x-y (horizontal unimplanted) planes indicate the cut-through sections displayed in subsequent figures.}
\label{F:F3_shape}
\end{figure}

Amplitude isosurfaces of the object from all five measured reflections are shown in Figure \ref{F:F3_shape}. Superimposed on this volume are the directions of Bragg vectors $\textbf{q}_\textit{hkl}$ for each of the five reflections. The morphology of the sample, recovered from different reflections, is consistent. Quantification of the spatial resolution was performed by differentiating line profiles in the x, y, z directions of the average electron density. By fitting a Gaussian profile to the air-crystal interfaces and taking the FWHM, the average spatial resolution was estimated to be \SI{28}{\nano\metre}. 

Upon recovery of the phase for each of the five reflections, visual inspection shows little indication of the presence of an implantation layer (see Supplementary Figure \ref{SF:SF1_phase}). The implantation layer is simply not visible when only the phase for each individual data set is considered. The reader should note that, as the sample was turned upside down during the fabrication process, the implanted layer will now appear at the bottom of the MBCDI object. Visibility of the implanted region is only marginally improved when one considers the strain in the direction of the scattering vector for each reflection (see Supplementary Figure \ref{SF:SF2_strain_hkl}). Upon careful inspection of the recovered strain, some indication of the implantation layer can be seen, visible as a layer of strain slightly higher than that of the bulk. The layer is slightly more visible for the (101) and (0$\bar{1}$1) reflections. 

\begin{figure}[H]
\centering\includegraphics[width=.5\textwidth]{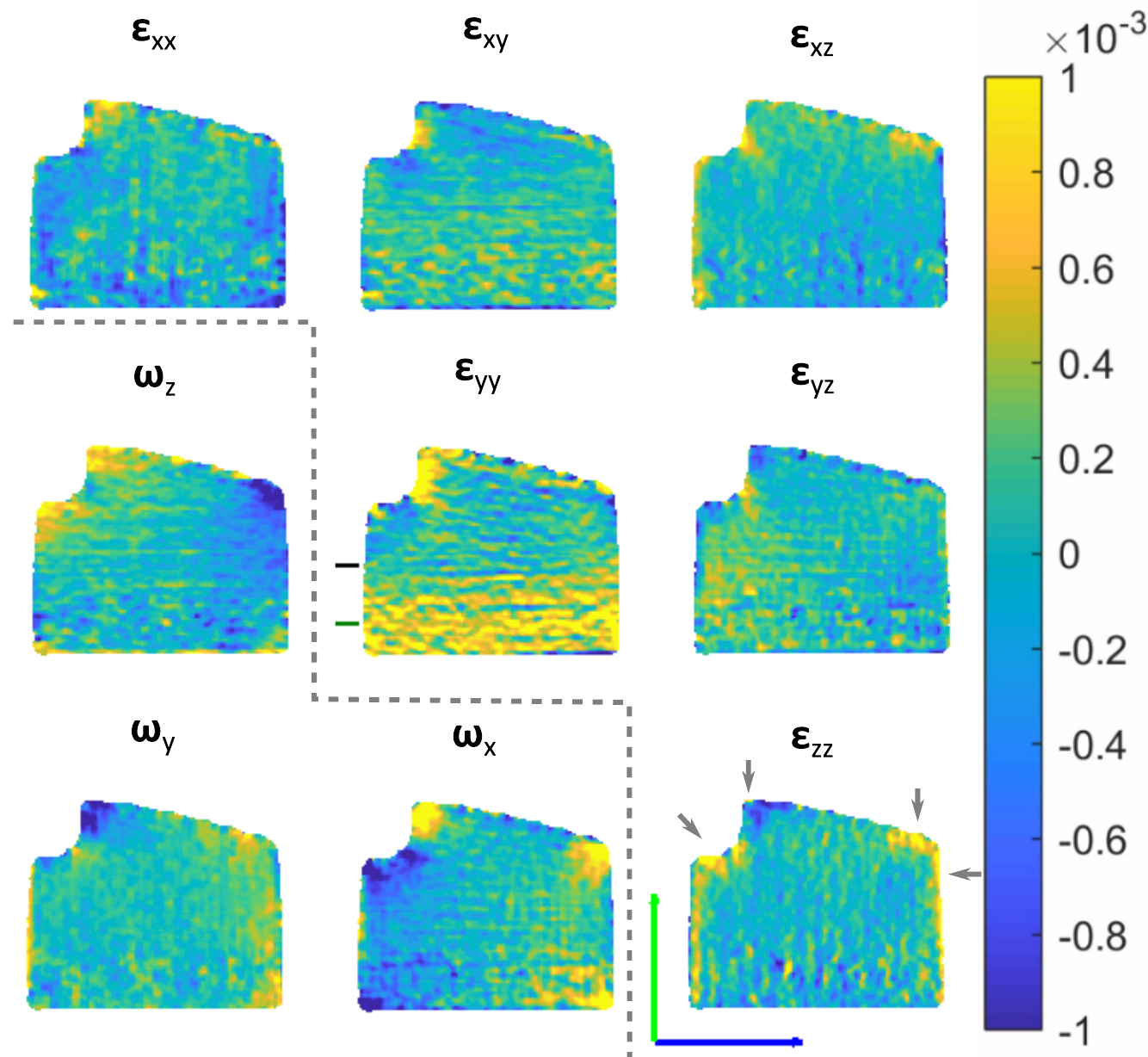}
\caption{Lattice strain tensor and lattice rotations recovered from MBCDI on the vertical section through the sample (red) shown in Figure \ref{F:F3_shape}, the other sections used in subsequent figures are shown by the dark green and black lines in the $\varepsilon_{yy}$ component. Shown are the six strain tensor components (upper right triangle and diagonal) alongside the three rotation tensor elements (bottom left triangle) as given in Table \ref{tab:strain_rot_tab}. The lattice swelling resulting from the tungsten self-ion implantation is most evident in the $\varepsilon_{yy}$ component. Residual effects from FIB fabrication are observable at the notch at the top left corners and the top right corner, indicated by the grey arrows. Heterogeneity of the strain within the implanted layer is indicative of a complex microstructure (particularly visible in the $\varepsilon_{xx}$, $\varepsilon_{yy}$ and $\varepsilon_{zz}$ components). The green and blue arrows are \SI{500}{\nano\metre} long. Lattice rotations are given in radians.}
\label{F:F4_StrainRotYZ}
\end{figure}

\begin{figure}[H]
\centering\includegraphics[width=1\textwidth]{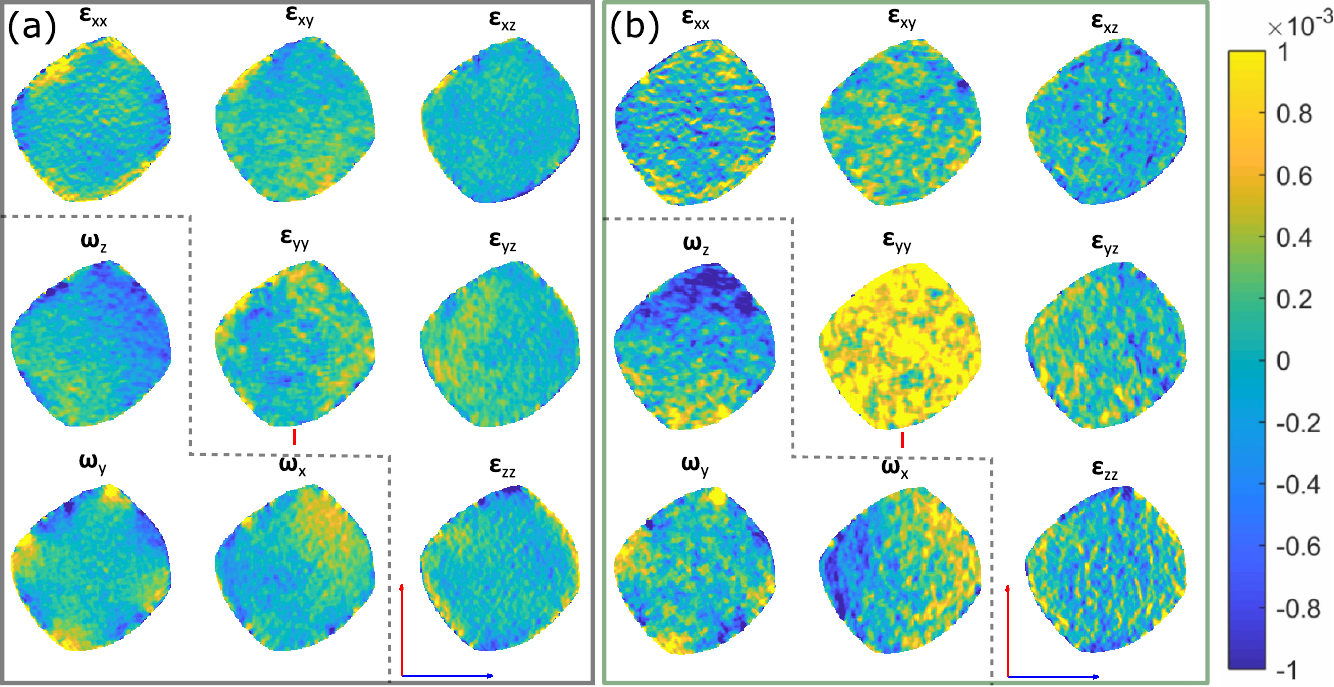}
\caption{Lattice strain tensor and lattice rotations recovered from MBCDI on (a) the horizontal unimplanted section (black) and (b) the horizontal implanted (green) section marked by the planes in Figure \ref{F:F3_shape}. Shown are the six strain tensor components (upper right triangle and diagonal) alongside the three rotation tensor elements (bottom left triangle) as given in Table \ref{tab:strain_rot_tab}. The lattice swelling resulting from the implantation is most evident in the $\varepsilon_{yy}$ component (the position of the vertical section (red) is also shown for this component). Heterogeneity of the strain within the implanted layer is indicative of a complex microstructure (particularly visible in the $\varepsilon_{xx}$, $\varepsilon_{yy}$ and $\varepsilon_{zz}$ components). The red and blue arrows are \SI{500}{\nano\metre} long. Lattice rotations are given in radians.}
\label{F:F5_StrainRotZX}
\end{figure}

Figure \ref{F:F4_StrainRotYZ} shows the recovered strain tensor, through the central slice of the sample. The implanted layer can be convincingly located in the $\varepsilon_{yy}$ component, which corresponds to the strain normal to the implanted surface (central component of Figure \ref{F:F4_StrainRotYZ}). The layer can be seen within the lower $1/3^{rd}$ of the sample and extends \SIrange[range-phrase = --, range-units = single]{200}{245}{\nano\metre} into the sample. The in-plane strain components are small, as was the case for He implanted W \cite{Hofmann2015b,Das2018,Das2018b}. The magnitude of the strain for the $\varepsilon_{yy}$ component averages \num{5e-4} and reaches a maximum of \num{2.9e-3} within the implanted layer. The strain sensitivity of the MBCDI measurement is of the order \num{2e-4} \cite{Hofmann2017,Hofmann2018,Hofmann2020}. Theoretically the strain sensitivity of MBCDI is limited by the numerical aperture to which data can be reliably phased. In practice this is affected by the stability of the experimental setup, the amount of coherent flux, imperfections such as chip-set boundaries in detectors or beamstops and the precision to which the geometry is known \cite{Hofmann2017,Hofmann2018,Hofmann2020,Carnis2019}. Addressing these factors in order to improve the strain resolution is an ongoing challenge within the BCDI community. Lattice expansion is again observed within the implanted layer. The measured thickness compares well to the simulated implantation profile, where the depth at which the dpa is half of its maximum was \SI{180}{\nano\metre}. The strain tensor and lattice rotation for horizontal slices are shown in Figure \ref{F:F5_StrainRotZX} for the unimplanted (a) and implanted (b) regions of the sample. The reader is encouraged to view Supplementary Movies \ref{SM:SM1_LatticeStrainRotatoionYZ} and \ref{SM:SM2_LatticeStrainRotatoionXZ} that show the strain tensor on a slice-by-slice basis through the entire thickness of the sample. 

Lattice swelling due to self-ion implantation damage is distinguishable from gallium (Ga)-FIB damage by the location within the strain microscopy sample and the thickness of the damage layer. The protective platinum cap, deposited before manufacture of the MBCDI sample, ensures that the lower surface is free of any Ga-FIB damage, whilst glancing angle Ga\textsuperscript{+} ion incidence and low energy polishing is used to minimise the effects at the FIB-milled surfaces of the sample volume. Even with careful fabrication it is not possible to completely remove all strain due to FIB damage. In this sample, FIB induced strain can be identified in the corners of the notched region where a combination of geometric shadowing and re-deposition reduces the effectiveness of low energy polishing. The remaining damage is confined to a thin surface layer, most visible at the top right corner. These features are marked by arrows in Figure \ref{F:F4_StrainRotYZ}. Overall, whilst present, the effects of FIB damage are small, readily identifiable, and can be excluded during further analysis of the data with respect to the tungsten-ion-damaged implantation layer.

\section{Discussion}
\label{S:discussion}
The in-plane strain components are close to zero throughout the recovered volume of MBCDI data, with the exception of some regions where FIB damage has not been completely removed. This is expected and agrees with reports in the literature for similar systems \cite{Hofmann2015b,Das2018,Das2018b}. Visualisation of the implanted layer is via an isosurface of the $\varepsilon_{yy}$ component, using a threshold of \num{1e-3} for positive lattice strain as shown in Figure \ref{F:F6_strain_iso} (a) (see animated version in Supplementary Movie \ref{SM:SM3_iso_movie}). To make a quantitative comparison between MBCDI and micro-beam Laue data, a strain profile as a function of depth was computed from the averaged strain in the core of the MBCDI sample. The average of the $\varepsilon_{yy}$ strain component for a \SI{200}{\nano\metre} diameter region occupying the \textit{x-z} plane is shown in Figure \ref{F:F6_strain_iso} (b). The top \SI{200}{\nano\metre} of the MBCDI sample (regions furthest from the implanted layer) have been excluded as they are more susceptible to the spurious FIB damage which is not entirely removed. Average $\varepsilon_{yy}$ strains of \num{3.5e-4} and \num{5e-4} are observed within the implanted layer for the \SI{20}{\mega\eV} (within \SI{1}{\micro\metre} of the surface, measured by EW-DAXM) and \SI{2}{\mega\eV} (average of the total implanted layer measured by MBCDI) implanted samples respectively. This agreement between implantation energies differing by an order of magnitude is quite remarkable. Importantly, this lends credibility to the use of higher ion energies for implantation when attempting to mimic neutron-based damage cascades. This allows for the generation of a sufficiently thick implantation layer for analysis techniques with limited depth resolution, including EW-DAXM, when the magnitude of the lattice strain is the primary concern.

\begin{figure}[H]
\centering\includegraphics[width=1\linewidth]{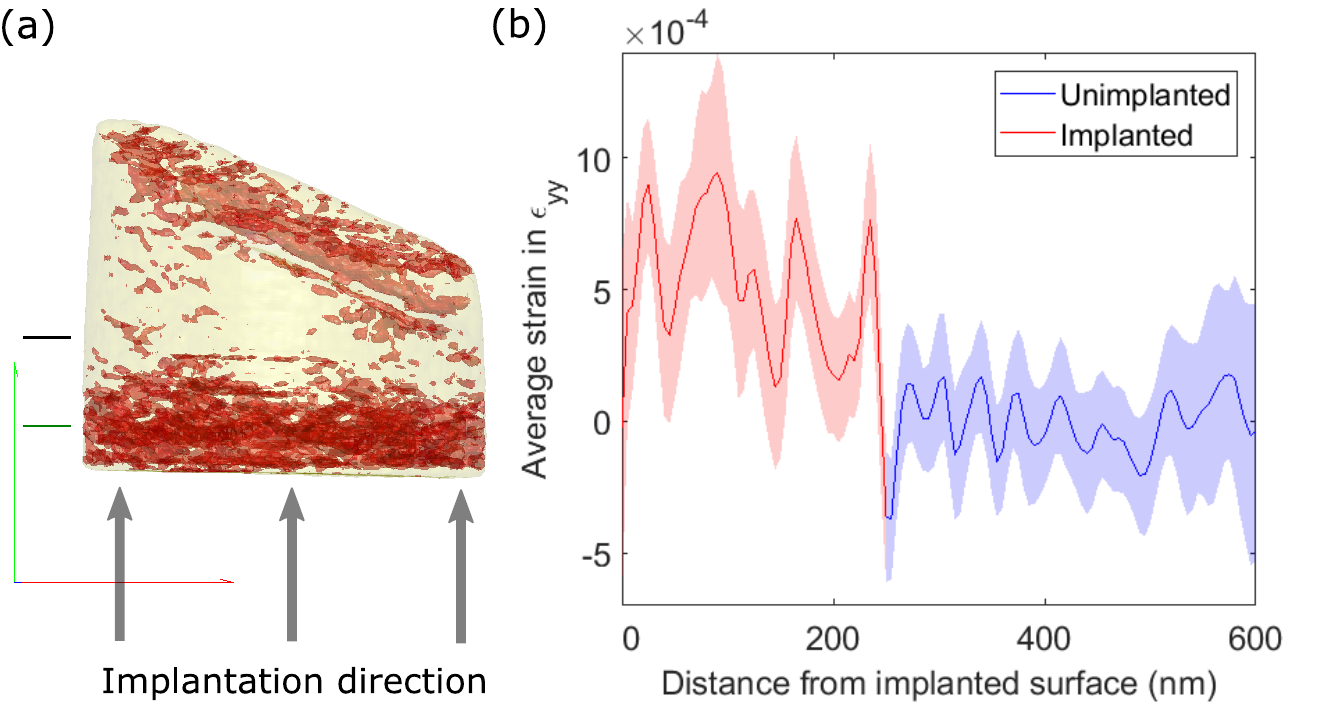}
\caption{(a) Rendering of the recovered strain microscopy sample showing a strain isosurface plotted for the $\varepsilon_{yy}$ component using a value of $1\times10^{-3}$. The position of the  horizontal sections in Figure \ref{F:F5_StrainRotZX} are shown by the black and green indicators. The red and green arrows are \SI{500}{\nano\metre} long. (b) Strain profile as a function of depth for the lower \SI{600}{\nano\metre} of the sample. The implanted layer is contained to the lower \SIrange[range-phrase = --, range-units = single]{200}{245}{\nano\metre}, noting that the implantation was performed into this lower surface and the sample flipped during fabrication in order to preserve this surface.}  
\label{F:F6_strain_iso}
\end{figure}

From the measured lattice strain tensor an estimate of the underlying defect population can be made. In tungsten vacancies are essentially immobile at room temperature, whilst SIAs are highly mobile even at cryogenic temperatures \cite{Nguyen-Manh2006,Ma2019}. The measured lattice swelling suggests that not all SIAs are lost to sample surfaces, but rather that a population of SIAs is retained within the damaged layer. The ratio of retained vacancies to SIAs cannot be unambiguously determined. However, we can make a lower bound estimate of the defect density, assuming retention of a population of Frenkel pairs (\textit{i.e.} that there are the same number of vacancies and SIAs). Hofmann \textit{et al.} showed that the out-of-plane lattice strain is related to the defect number density by:

\begin{equation}\label{eq:4}
    \varepsilon_{yy}=\frac{1}{3}\frac{(1+\nu)}{(1-\nu)}\sum_{A}n^{(A)}\Omega_r^{(A)},
\end{equation}

where $A$ is the defect type, and the Poisson ratio, $\nu = 0.28$, of tungsten is used \cite{Hofmann2015b}. Given the relaxation volumes of vacancies and SIAs from Hofmann \textit{et al.} \cite{Hofmann2015b}, the number density of Frenkel pairs in the implanted layer for both implantation conditions can be estimated as $\sim{400}$ appm. It is important to bear in mind that this analysis neglects the effect of SIA clustering on the relaxation volume, though for interstitial loops containing up to several hundred SIAs it has been shown that the relaxation volume scales almost linearly with the number of SIAs in the loop \cite{Hofmann2015b}. The number density of Frenkel pairs estimated from X-ray diffraction data is significantly greater than observed by TEM, where the number of defects is typically reported to be in the range of \SIrange[range-phrase = --, range-units = single]{10}{250} appm \cite{Tanno2011,Yi2015,Yi2016}. This is because small defects with a size below \SI{1.5}{\nano\metre} remain invisible when measured via TEM \cite{Zhou2006}, whilst X-ray diffraction based measurements are sensitive to the contribution from all defects. 

\begin{figure}[H]
\centering\includegraphics[width=1\linewidth]{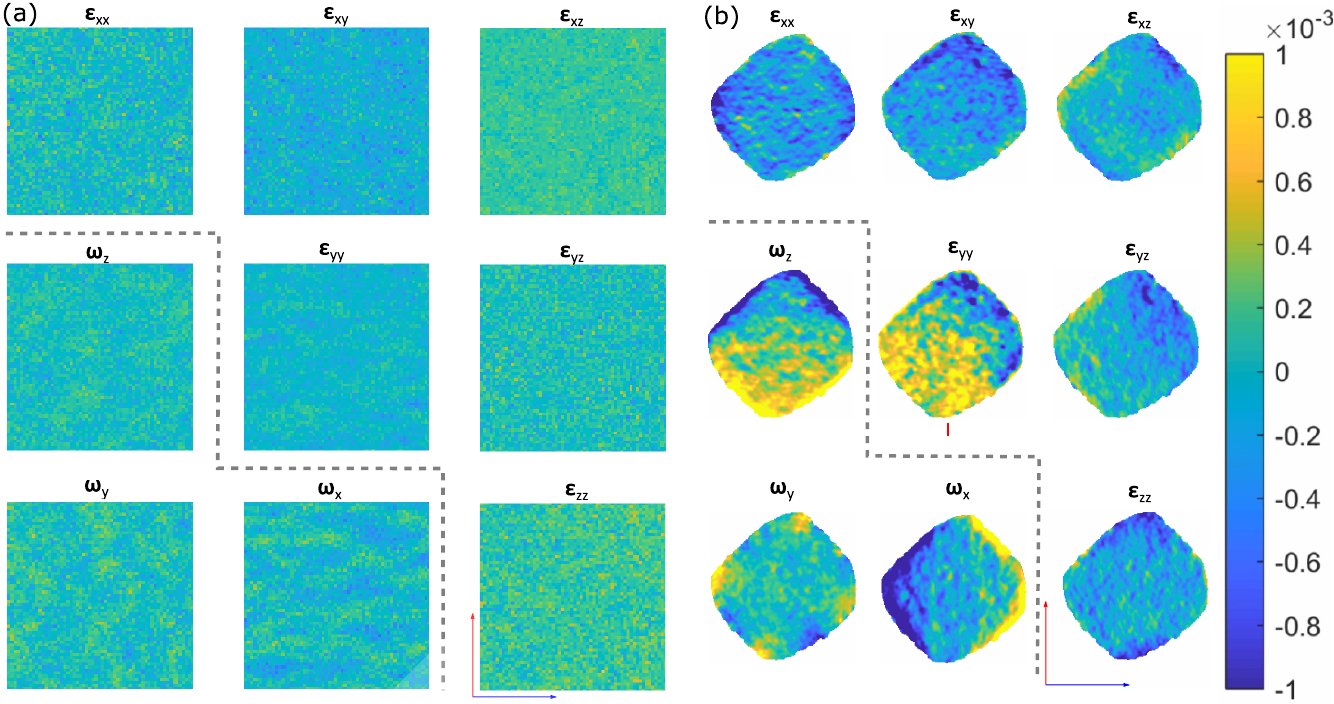}
\caption{Lattice strain tensor and lattice rotations recovered from HR-EBSD (a) of \SI{2}{\mega\eV} self-ion implanted tungsten captures the fine heterogeneity of strain at the surface of the implanted layer indicating non-uniformity of the damage. Lattice strain tensor and lattice rotations recovered from MBCDI and used to simulate HR-EBSD data (b) by averaging the first two voxels (\SI{10}{\nano\metre}) of the implanted layer. This signal volume mimics the signal expected from HR-EBSD measurement of the same sample, where the depth-contribution reaches \SI{50}{\percent} at \SI{9}{\nano\metre} from the surface when using an accelerating voltage of \SI{20}{\kilo\electronvolt} and a sample tilt of \SI{70}{\degree}. Note that the pixel size in the transverse direction has been down-sampled to \SI{15}{\nano\metre} to better represent a realistic pixel size for HR-EBSD. Shown are the six strain tensor components (upper right triangle and diagonal) alongside the three rotation tensor elements (bottom left triangle) as given in Figure \ref{tab:strain_rot_tab}. A reduction in the magnitude of strain is seen when compared to regions deeper within the implanted layer (see Figure \ref{F:F5_StrainRotZX}(b)). The red and blue arrows are \SI{500}{\nano\metre} long. Lattice rotations are given in radians.}
\label{F:F7_HREBSDvsBCDIab}
\end{figure}

Interestingly, closer inspection of the strain components obtained from the MBCDI measurement show clear evidence of spatial heterogeneity. This is particularly evident in the $\varepsilon_{xx}$, $\varepsilon_{yy}$ and $\varepsilon_{zz}$ components of Figure \ref{F:F4_StrainRotYZ} and \ref{F:F5_StrainRotZX}(b). The mottled features have a size in the range of \SIrange[range-phrase = --, range-units = single]{20}{70}{\nano\metre}, in some cases forming longer, chain-like features. Independent simulations of irradiation damage evolution have concurrently predicted that spatially fluctuating stress fields within an evolving cascade microstructure will influence the behaviour of subsequent cascades, resulting in the formation of complex damage microstructure with long range order evidenced as strong variations in stress and strain \cite{Derlet2020}. The measurements of strain heterogeneity presented here complement two-dimensional TEM observations of damage microstructure and provide, in three-dimensions, experimental verification of the predicted defect ordering at length scales considerably larger than those accessible via atomistic simulation. The observed defect structure is similar to that reported in the TEM study by Ciupinski \textit{et al.} \cite{Ciupinski2013}. Interestingly, both the magnitude of strain and the spatial heterogeneity are reduced near the surface of the implanted layer for MBCDI data. This is most clearly captured in the simulated HR-EBSD map for the BCDI sample, in which only the top two voxels (\SI{10}{\nano\metre}) adjacent to the sample surface are considered  (see Figure \ref{F:F7_HREBSDvsBCDIab}(b)). As the strain is indicative of the defect content, a reduced strain implies a reduction of defects near this free surface. This is in line with previous observations of irradiation damage near free surfaces, where the surface acts as a significant defect sink in the first \SIrange[range-phrase = --, range-units = single]{10}{15}{\nano\metre} \cite{Yu2018,Mason2014}. 

The microstructure is qualitatively similar to the non-uniformity observed for other grains within the same sample measured via HR-EBSD, shown in Figure \ref{F:F7_HREBSDvsBCDIab}(a). Here the feature size is in the range of \SIrange[range-phrase = --, range-units = single]{40}{120}{\nano\metre}. When comparing MBCDI measurements with HR-EBSD, we note that the magnitude and heterogeneity are smaller in the HR-EBSD measurements. This may be attributed to free surface effects that can have a significant influence on HR-EBSD observations, which are sensitive to the deviatoric strain in the first few nanometers of material \cite{Winkelmann2010,Hardin2015}. Calculations from 100,000 electron trajectories performed by Monte Carlo Simulation of Electron Trajectory in Solids (Casino) \cite{Drouin2007} show that \SI{50}{\percent} of the detected electrons in the HR-EBSD data presented here originate from the top \SI{9}{\nano\metre} of the material, a distance over which some defects may escape to the free surface. The difference in observations for HR-EBSD and TEM, with respect to MBCDI and micro-beam Laue diffraction,  highlights that EM methods may only capture the smaller strains and strain variations at the sample surface. As defects escape to nearby free surfaces and TEM fails to account for the invisible defect population, these factors are thought to contribute to the significantly lower reported defect populations from EM methods. MBCDI, on the other hand, shows that strains beneath the surface of the sample can be significantly larger. This enforces the merit of adopting three-dimensionally resolved techniques when considering strain at the nano-scale. 

For implantation doses below 0.1 dpa, heterogeneity in strain components is no longer visible via HR-EBSD. Here, we attribute this to defects being randomly distributed at low doses when there is little cascade overlap. This random defect population results in similar average strain from the probed volume across the sample. However, as the dpa increases, so too does the proximity of cascades with respect to one and-other. We hypothesise that this results in the new cascades being influenced by the history within the local lattice, i.e. the pre-existing defects and associated  strain fields. A detailed study by Yu \textit{et al.} \cite{Yu2020} using electron channelling contrast imaging (ECCI) and HR-EBSD further supports this. These findings are in line with the simulation findings of Ma \textit{et. al} \cite{Ma2019}, which show that the local lattice stress has a substantial effect on the evolution of defects. This has also been observed by TEM in self-ion implantation of tungsten \cite{Yi2016,Yi2015}. Whilst cascade overlap is initially a stochastic process, long range order develops as an increasing number of cascade events occur and become sufficiently close so as to interact with the stress field from defects associated with previous cascades. This results in the formation of the heterogeneous microstructure observed in both the MBCDI and HR-EBSD measurements. Damage cascade formation, evolution and eventual spatial heterogeneity may be affected by the ion energy, especially as the energy approaches the sub-cascade splitting energy. Further, as the implantation thickness decreases, the cascade volume may become insufficient to fully accommodate the evolving microstructure which occurs at higher implantation energies or when considering neutron irradiation. This complicated regime remains largely unexplored. Whether higher implantation energy increases the heterogeneity of the damage structure due to increased cascade splitting and overlapping of sub-cascades needs to be studied in more detail in the future.

\section{Conclusions}
\label{S:conclusions}
In this work we have gained insight into damage processes in tungsten caused by self-ion implantation at \SI{20}{\mega\eV} and \SI{2}{\mega\eV}. The thirty-fold improvement in resolution (compared to X-ray micro-beam Laue diffraction) afforded by MBCDI enables non-invasive measurement of the strain tensor associated with the damaged material in three dimensions. This makes depth resolved strain measurements of samples implanted at significantly lower ion energies possible. Consequently, this reduces the potential difference in damage microstructure due to differences in PKA energy between self-ions and fusion neutrons. We have found that:
\begin{enumerate}
\item The magnitude of out-of-plane strain is comparable between \SI{20}{\mega\eV} and \SI{2}{\mega\eV} implantation energies (\num{3.5e-4} and \num{5e-4} respectively). This suggests that the average lattice strain is insensitive to the energy of the incident ions.
\item Lattice swelling suggests retention of certainly some population of SIAs and a lower bound estimate of this population can be made (400 appm, assuming retention of Frenkel pairs). Interestingly this estimate is substantially larger than previous estimates from TEM observations, suggesting that a large number of small point defects remain invisible to TEM.
\item Substantial strain heterogeneity is observed in the ion-implanted layer. This suggests a clustering of defects and the formation of larger structures. At the surface this clustering is also seen by HR-EBSD. It is hoped that these results can be used to inform future models so that the effects of spatial heterogeneity in the microstructure of irradiated samples might be better accounted for.
\item Strains are smaller near the surface than in the bulk, indicating that there is a loss of defects to the free surface. Care needs to be taken when using surface sensitive techniques to probe strain and microstructure due to ion-implantation as these may not be wholly representative. 
\end{enumerate}   

\section*{Acknowledgements}
\label{S:acknowledgements}
NWP, DY and FH acknowledge funding from the European Research Council (ERC) under the European Union's Horizon 2020 research and innovation programme (grant agreement No 714697). HY and SD acknowledge support from The Leverhulme Trust under grant RPG-2016-190. The authors acknowledge use of characterisation facilities within the David Cockayne Centre for Electron Microscopy, Department of Materials, University of Oxford, the use of the University of Oxford Advanced Research Computing (ARC) facility http://dx.doi.org/10.5281/zenodo.22558 and E. Tarlton for the use of computational resources. The Zeiss Crossbeam FIB/FEG SEM used in this work was supported by EPSRC through the Strategic Equipment Fund, grant \#EP/N010868/1. This work was supported by EUROfusion Enabling Research project TRiCEM, Tritium Retention in Controlled and Evolving Microstructure. Ion implantations were performed at the Helsinki Accelerator laboratory, Department of Physics, University of Helsinki. Diffraction experiments used the Advanced Photon Source, a US Department of Energy (DOE) Office of Science User Facility operated for the DOE Office of Science by Argonne National Laboratory under Contract No. DE-AC02-06CH11357.





\bibliographystyle{model1-num-names}
\bibliography{ArxivV1.bbl}
\biboptions{sort&compress}






\label{SI:phasing}
\label{SF:SF1_phase}
\label{SF:SF2_strain_hkl}
\label{SM:SM1_LatticeStrainRotatoionYZ}
\label{SM:SM2_LatticeStrainRotatoionXZ}
\label{SM:SM3_iso_movie}

\end{document}